\documentclass[%
 reprint,
 amsmath,amssymb,
 aps,
]{revtex4-1}

\newcommand{\MetastableState}{2^{3\!}S_1}%
\newcommand{\UpperS}{5^{3\!}S_{1}}%
\newcommand{\PStateManifold}{2^{3\!}P}%
\newcommand{\UpperStateManifold}{5^{3\!}D}%
\newcommand{\triplet}{^{3\!}}
\newcommand{\singlet}{^{1\!}}
\usepackage{gensymb}
\usepackage{svg}
\usepackage{graphicx}
\usepackage{dcolumn}
\usepackage{bm}
\usepackage{hyperref}

\begin{document}


\title{
Frequency measurements of transitions from the $2\triplet P_2$ state \\to the $5\singlet D_2$ $5\triplet S_1$, and $5\triplet D$ states in ultracold helium}

\author{Jacob A. Ross}
\affiliation{Laser Physics Centre, Research School of Physics, The Australian National University\\}
\author{Kieran F. Thomas}
\affiliation{Laser Physics Centre, Research School of Physics, The Australian National University\\}
\author{Bryce M. Henson}
\affiliation{Laser Physics Centre, Research School of Physics, The Australian National University\\}
\author{Danny Cocks}
\affiliation{Laser Physics Centre, Research School of Physics, The Australian National University\\}
\author{Kenneth G. H. Baldwin}
\affiliation{Laser Physics Centre, Research School of Physics, The Australian National University\\}
\author{Sean S. Hodgman}
\affiliation{Laser Physics Centre, Research School of Physics, The Australian National University\\}
\author{Andrew G. Truscott}
\affiliation{Laser Physics Centre, Research School of Physics, The Australian National University\\}

%



\begin{abstract}
We perform laser absorption spectroscopy with ultracold $^4$He atoms to measure the energy intervals between the $2\triplet P_2$ level and five levels in the n = 5 manifold. The laser light perturbs the cold atomic cloud during the production of Bose-Einstein condensates and decreases the phase space density, causing a measurable decrease in the number of atoms in the final condensate. We improve on the precision of previous measurements by at least an order of magnitude, and report the first observation of the spin-forbidden $2^{3\!}P_2 - 5^{1\!}D_2$ transition in helium. Theoretical transition energies agree with the observed values within our experimental uncertainty.
\end{abstract}

\maketitle


\section{Introduction}

  The appearance of ordinary matter arises from interactions between charged particles and light. This phenomenon is the domain of the theory of quantum electrodynamics (QED), which provides the most accurate quantitative predictions of any physical theory to date. The theory of QED is the workhorse of modern atomic structure calculations, whose only inputs are the CODATA values of three physical constants: the proton-electron mass ratio, the Rydberg constant, and the fine structure constant $\alpha$. These constants of nature can be constrained with state-of-the-art atomic spectroscopy, which is accurate enough to match theoretical uncertainties in table-top experiments. Thanks to the quality of modern theory and experiment, atomic structure measurements reprise their role in frontier tests of physics. 

  In 1964 Schwartz proposed the determination of $\alpha$ from the fine structure intervals of the $2\triplet P$ manifold in helium \cite{Schwartz64}, which are subject to strong QED effects. The contemporary knowledge of helium's structure greatly exceeds Schwartz's anticipation of parts-per-million accuracy. For example, the $2\triplet S_1 - 2\triplet P$ and $2\triplet P - 3\triplet D$ intervals measured by Cancio Pastor \textit{et al.} \cite{Pastor04} and Luo \emph{et al.} \cite{Luo16}, respectively, both have relative uncertainties better than 50 parts per \emph{trillion}, providing Lamb shift measurements accurate to several ppm. The measurement of the $\PStateManifold$ fine structure splitting by Smiciklas \textit{et al.} to sub-kilohertz precision determines $\alpha$ to several ppb \cite{Smiciklas10}. Measurements of the $2^{3\!}P_1-2^{3\!}P_2$ interval by Kato \emph{et al.}, accurate to 25Hz, would constrain $\alpha$ to less than one ppb given a similarly accurate measurement of the $2^{3\!}P_0 - 2^{3\!}P_1$ transition and QED calculations including terms of order $\alpha^7$ \cite{Kato18}. 

  \begin{figure}[b]
      \centering
      \includegraphics[width=0.45\textwidth]{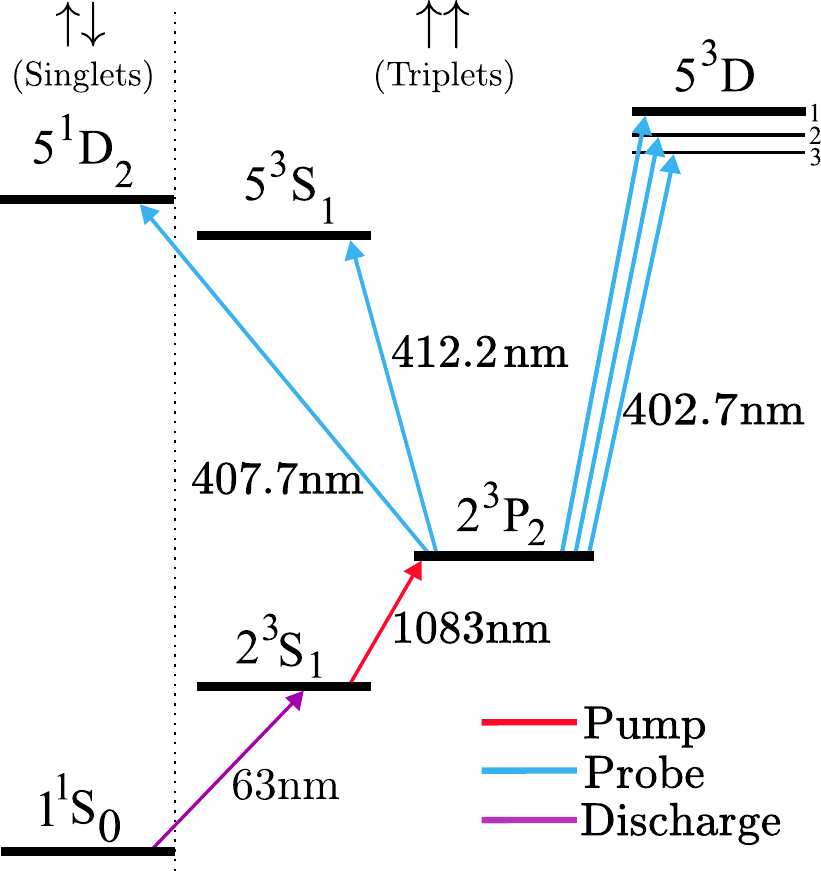}
      \caption{Energy level diagram for the helium atom. The transitions measured in this work (blue) are driven by a tunable laser referred to in the text as the \emph{probe beam}. A laser tuned to the $2\triplet S_1-2\triplet P_2$ transition (red, referred to as \emph{pump beam}) populates the lower state of the target transitions.  The doubly forbidden $1^{1\!}S_0 - 2^{3\!}S_1$ transition is excited in a high voltage discharge source. Transitions across the dotted line are \emph{forbidden} by the $\Delta S=0$ selection rule. Level splittings are not to scale.}
      \label{fig:lvl_diag}
  \end{figure}

  A concurrent issue is the so-called `proton radius puzzle': Determinations of the proton charge radius from Lamb shift measurements in muonic and electronic hydrogen \cite{Pohl10, Bezginov19}, electron-hydrogen scattering experiments \cite{Beyer17,Xiong19}, and isotope shifts in light muonic atoms \cite{Kalinowski19,Pohl16} disagree significantly with both the CODATA recommended value and with other recent experiments \cite{Fleurbaey18}.  Helium is a promising candidate to provide insight into this unresolved issue because its simple structure is tractable to QED calculations. Ongoing theoretical work \cite{Pachucki15,Pachucki17,Pachucki11,Pachucki10,Morton12,Morton06,Patkos16,Patkos17} and recent high-precision measurements \cite{Rooij11,Notermans14,Notermans16,Rengelink18} find a $4\sigma$ discrepancy between the difference $\delta = r^2(^3\textrm{He}) - r^2(^4\textrm{He})$ of squared nuclear charge radii obtained from the isotope shifts of the $\MetastableState - \PStateManifold$ and ${2^{1\!}S_{0}} - \MetastableState$ transitions \cite{Pachucki15,Patkos17}. The completed calculation of QED effects to order $\alpha^7$ will allow determination of the absolute nuclear charge radii accurate to better than 1\% \cite{Pachucki17}. Along with these $\alpha^7$ contributions, measurement of the $2\triplet P-2\triplet S$ spacing to within 1.4kHz would allow a determination of the nuclear charge radius to below 0.1\% accuracy, better than expected from the muonic helium Lamb shift \cite{Wienczek19}.  

  Notable among recent studies of helium's structure are the measurements of \emph{forbidden} transitions between the singlet and triplet manifolds. Such transitions are made possible in reality due to relativistic effects and are extremely narrow, therefore precise measurements of their spectral features can provide stringent tests of QED \cite{Lach01}. The present work complements existing measurements of forbidden lines in helium at 1557nm \cite{Rooij11,Rengelink18}, 887nm \cite{Notermans14}, and 427nm \cite{Thomas20}.

  In this work, we report on frequency measurements of the transitions from the $2^3P_2$ state to five states in the $n=5$ manifold of $^4$He, illustrated in Fig. \ref{fig:lvl_diag}. 
  Our results improve on the precision of past measurements \cite{Martin60} by at least an order of magnitude.
  We resolve the fine structure splitting of the $\PStateManifold_2 - 5\triplet D$ transition. As far as we can ascertain, our work includes the first observation of the spin-forbidden $2^3P_2 - 5^1D_2$ transition in helium, whose transition rate is four orders of magnitude smaller than the other transitions reported here.

\section{Experiment}

We performed our measurements by disrupting a laser cooling stage in the production of a Bose-Einstein condensate (BEC) of $^{4}\textrm{He}$ atoms. Our experimental sequence begins with $\sim10^8$ atoms in the metastable $2\triplet S_1$ state, cooled to  $\sim1 \textrm{mK}$ in a magneto-optical trap. Throughout the sequence, all laser cooling light is tuned to the $2\triplet S_1 - 2\triplet P_2$ transition at 1083.331nm \cite{Shin16}. The light is then switched off, leaving only the $m_J=1$ atoms in a magnetic trap generated by field coils in a BiQUIC configuration \cite{Dall07}. Next, during a Doppler cooling stage, we illuminated the atoms with $\sim$30$\mu$W$/m^2$ of $\sigma^+$ polarized cooling light in an approximately uniform magnetic field set by the high bias of the BiQUIC coils, further cooling the atoms to $\sim200\mu \textrm{K}$. Finally, we applied forced evaporative cooling by RF radiation, cooling the sample below the critical temperature to form a BEC. Each iteration of this procedure produced a BEC of $\sim 5\times10^5$ atoms in a cigar-shaped harmonic trap with trapping frequencies $\omega = 2\pi (425,425,45)$ Hz. 

At the end of the experimental sequence the atoms are in the metastable $2^3S_1(m_J=1)$ state, which has a lifetime of $7870$ seconds \cite{Hodgman09}. The metastable state is 19.8eV above the true ground state, and this large internal energy enables single-atom detection by a multi-channel plate and delay-line detector stack \cite{Manning10}, located $\sim$850mm below the trap. The detector has a quantum efficiency of $\sim 8\%$ and saturates at high atom fluxes, precluding accurate number measurements by simply dropping a BEC on the detector. Instead, we used a \emph{pulsed atom laser}, wherein  broadband radio-frequency pulses transfer $\sim$2\% of the trap population to the untrapped $2\triplet S_1(m_J=0)$ state \cite{Manning10,Henson18}. The resulting coherent matter-wave pulse falls onto the detector, allowing the atom number and temperature to be accurately determined without saturating the detector. Each iteration of the BEC preparation sequence followed by detection is referred to as a \emph{shot}. Our data collection protocol consisted of a cycle of one calibration shot with the probe beam switched off, followed by one measurement shot at each of two magnetic field strengths with the probe beam on. 

The physical basis of our measurement is the sensitivity of forced evaporative cooling to the initial conditions of the helium atoms. The precise effect of photon scattering on the final cloud properties depends on the exact details of the evaporation sequence, and is hard to model exactly. Instead, we give a qualitative picture of the role evaporative cooling plays in transforming photon scattering to a measurable change in trap population. 

During the Doppler cooling stage of BEC creation, the 1083nm cooling beam acts as an optical pump and excites atoms to the $2\triplet P_2(m_J=2)$ state. From the $2\triplet P_2$ state they may decay, with a $\sim$97ns lifetime, back to the trapped metastable state or absorb photons from the probe beam and become excited again to the target state. Doubly-excited atoms may decay back to the trapped $m_J=+1$ state of the $2\triplet S_1$ level, in which case the photon absorption and emission events add heat to the cloud by imparting a nonzero average impulse to the atoms. An initially hotter cloud therefore reduces the efficiency of evaporative cooling, resulting in a higher atom loss during the process and a lower final number in the trap. 

Alternatively, atoms may decay to other untrapped magnetic states of the metastable state, or to the true ground state via a spin-flip transition to the singlet manifold. Decay to untrapped states reduces the initial atom number and can even impart heat to the cloud as these atoms leave the trap - via scattering with trapped atoms. This heating will be much smaller than in the previous case because the scattering rate will be small in such a dilute gas. However, reducing the initial trap population also manifests as a reduced final atom number. In both cases a photon scattering signal clearly manifests in the reduction of the total trapped final number $N$ relative to the final number $N_c$ in calibration shots. We define our signal to be the relative number loss $(N_c-N)/N_c$ to compensate for drift in the trap population over time.

\begin{figure}[t]
    \centering
    \includegraphics[width=0.5\textwidth]{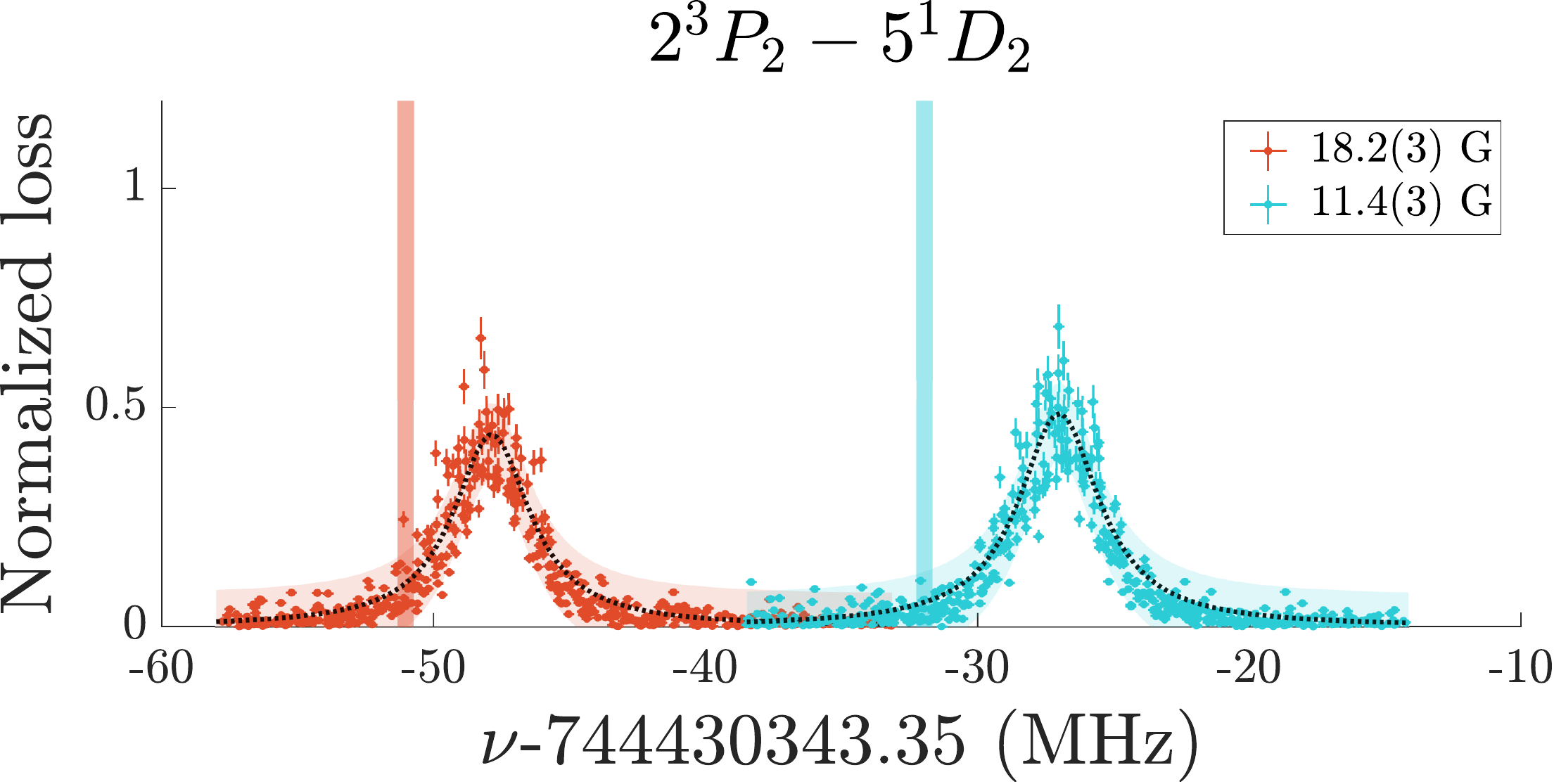}
    \caption{Line profile for the spin-forbidden $\PStateManifold_2 -  5^{1\!}D_2$ transition, showing normalized atom number loss versus probe laser frequency $\nu$, as measured in an 18.2(3)G (red) and 11.4(3)G (blue) background field, with Lorentzian fits (black dotted line, with prediction confidence interval shaded). Error bars account for detector efficiency and calibration model uncertainty. For comparison, theoretical predictions (vertical bars) Zeeman shifted from the predicted zero-field value \cite{Drake07} according to the field calibration, whose uncertainty (shaded width) is dominated by background field measurements.}
    \label{fig:1D_2_line}
  \end{figure}

To generate the probe beam light, we used 532nm light from a Lighthouse Photonics Sprout module to pump a tunable M-squared SolsTi:S titanium-sapphire laser, tuned near 800nm, and frequency-doubled the output in an M-squared ECD-X module to produce the target wavelengths. A sample of the Ti:S output was fibre coupled to a High Finesse WS8 wavemeter. A software lock used the wavemeter output to stabilize the laser and to scan the probe beam frequency across the region of interest. We calibrated the wavemeter by saturated absorption spectroscopy of cesium in a vapor cell. Our frequency reference was the $6S_{1/2}(F=3)-8S_{1/2}(F=3)$ two-photon transition in cesium at $364507238.36(1)$MHz \cite{Wu13}. To minimize wavemeter drifts over time, we calibrated the wavemeter daily, observing maximum drifts of order 1MHz over this timescale. 

We used the first diffracted mode of an acousto-optical modulator (AOM) to control the probe beam power. The output of the AOM was fibre coupled to the vacuum insertion optics, where a photodiode provided the input for PID control of the beam power. The profile and polarization of the beam were set manually with lenses and waveplates prior to vacuum entry. Additional details about the experimental setup can be found in the supplementary materials of Ref. \cite{Thomas20}.

\begin{figure}[t]
    \centering
    \includegraphics[width=0.48\textwidth]{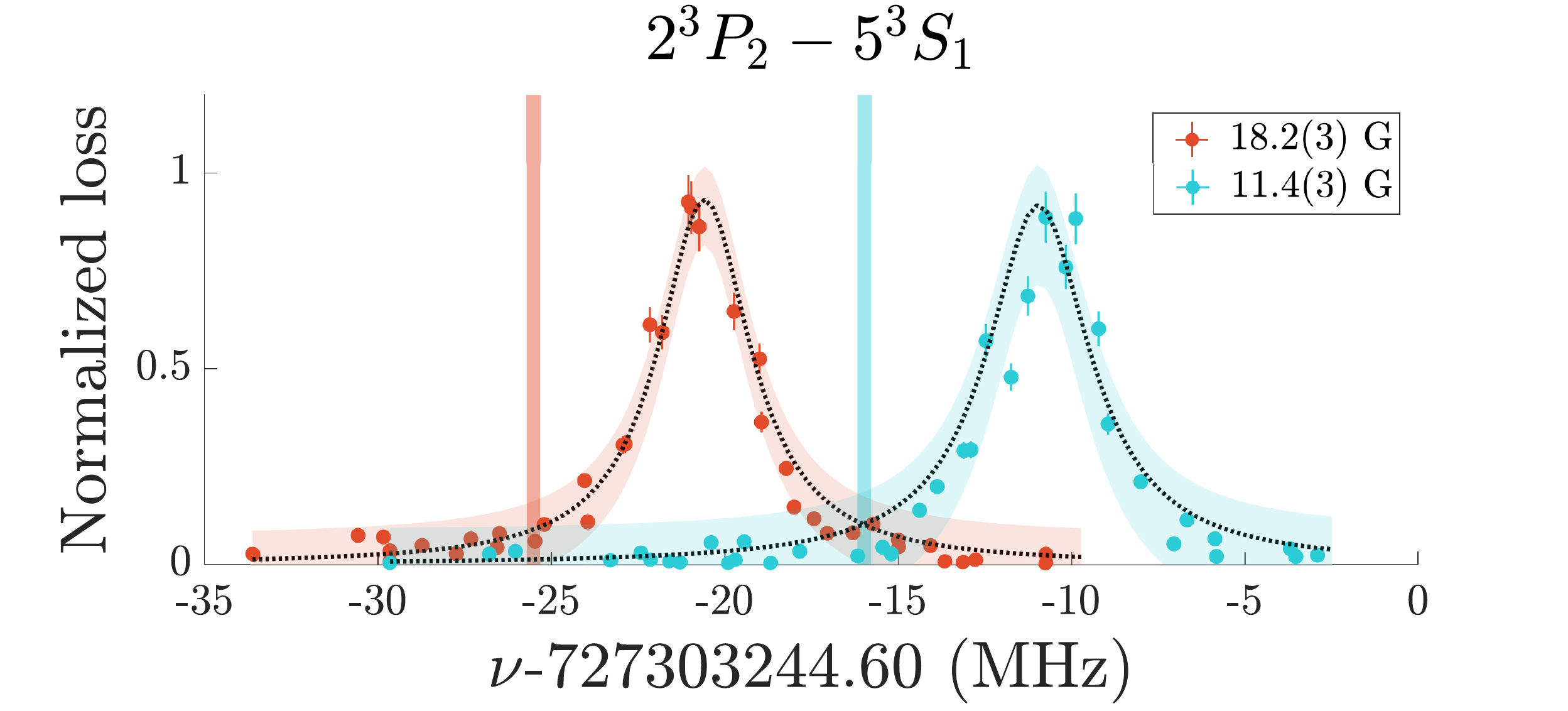}
    \caption{Measured atom loss versus laser frequency for the $2\triplet P_2 - 5\triplet S_1$ resonance in comparison to Zeeman-shifted predictions \cite{Drake07}, shown as for Fig. \ref{fig:1D_2_line}}
    \label{fig:5_3S_1}
\end{figure}

To make our measurements of the transition frequencies for the $5\singlet D_2$ and $5\triplet S_1$ states, we illuminated the atoms with the probe light for periods of order 100 ms during the Doppler cooling stage. The exact duration differed for each line to obtain a good SNR without saturating the atom loss. The light was $\sigma^-$ polarized to drive transitions to the $m_J=1$ states of the upper levels. For the forbidden $5\singlet D_2$ transition the beam was focused on the atom cloud with a waist of approximately $100\mu$m and a peak intensity of order $5\times 10^3$ W/m$^2$. For all other measurements the beam was collimated with a peak intensity of order $ 5$ W/m$^2$. We took measurements at two points in the Doppler cooling stage with a bias field strengths of 18.2(3) and 11.4(3) Gauss, which we calibrated independently by RF spectroscopy. For each field strength we obtained the atom loss (with respect to calibration shots) versus probe laser frequency. After correcting for the measured AOM and vapor cell shifts, we fit the measured response with a Lorentzian function plus constant background, as shown in Figs. \ref{fig:1D_2_line} and \ref{fig:5_3S_1}.

We correct for the linear Zeeman shift using the measured field strengths to estimate the field-free transition frequencies with sub-MHz statistical uncertainty. This determines the $2\triplet P_2-5\singlet D_2$ and $2\triplet P_2-5\triplet S_1$ transition energies to be 3MHz and 5MHz larger, respectively, than the predictions presented in \cite{Drake07}. However, the absolute accuracy of these measurements is limited by our instrumentation. Our results (Tab. \ref{tab:results}) are consistent with current predictions \cite{Drake07} within $2\sigma$ after accounting for all systematic uncertainties (Tab. \ref{tab:errors}).

\section{5$\triplet$D fine structure}

\begin{figure}[t]
  \includegraphics[width=0.5\textwidth]{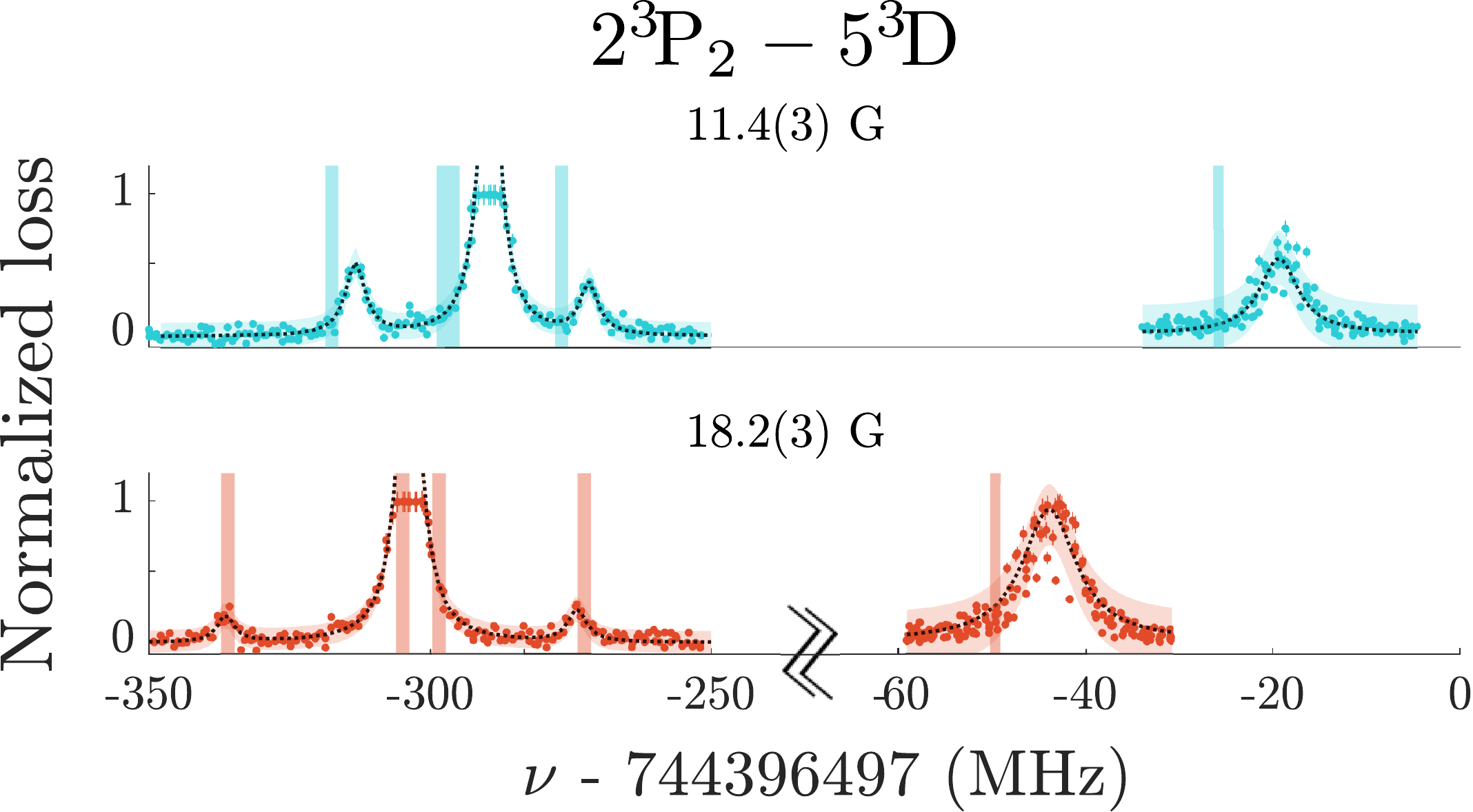}
  \caption{Line profiles for the $\PStateManifold_2 -  5^{3\!}D$ transitions, shown as for Figs. \ref{fig:1D_2_line} and \ref{fig:5_3S_1}. The normalized loss is shown versus probe laser frequency for the two different field strengths with a common horizontal scale. Theory lines indicate predictions from \cite{Drake07} after applying the relevant Zeeman shift. N.B. the scale break here and in Fig. \ref{fig:fitting_3D} coincide.}
  \label{fig:combined_5D_lines}
\end{figure}

Unlike the $5\singlet D_2$ and $5\triplet S_1$ levels, the $5\triplet D$ manifold splits into fine structure sublevels, leading to multiple absorption peaks and requiring a more involved analysis. We drove transitions to the $5\triplet D_J, J\in\{1,2,3\}$ levels with a combination of $\pi$ and $\sigma^-$ polarized light (in the atomic frame) and obtained four peaks, as shown in Fig. \ref{fig:combined_5D_lines}. The saturated peak near -300MHz (relative to the predicted $2\triplet P_2 - 5\triplet D_1$ interval) is in fact two peaks corresponding to the $5\triplet D_2(m_J=1)$ and $5\triplet D_3(m_J=2)$ states, which are separated by less than their linewidth. This illustrates a shortcoming of our technique, namely the limited dynamic range. For measurements of single peaks this is not an issue as the total irradiated energy can be adjusted to obtain a good singal-to-noise ratio without completely depleting the BEC. In this case, however, there is a trade-off between keeping the small peaks above the noise floor and preventing the superposed peaks from saturating. This limitation could be eased with a larger initial condensate because the dynamic range is essentially limited by the atom loss.

The Zeeman shift of the $J=2$ and $J=3$ levels is comparable to the interval between them, and so the mixing of levels means the correction is no longer proportional to $B$. Instead, we solve the eigenvalue optimization problem 
\begin{equation}
\textrm{min}_{E_{\textrm{fs}}} \sum_{J,m_J} \left(\nu_{{J,m_J}}^{\textrm{{pred}}}(E_{\textrm{fs}},B) - \nu_{{J,m_J,B}}^{\textrm{{obs}}}\right)^2,
\label{eqn:opt-problem}
\end{equation}
which minimizes the squared error between observed and predicted transition frequencies ($\nu^{\textrm{{obs}}}$ and $\nu^{{\textrm{pred}}}$ respectively), summed over all relevant $|J,m_J\rangle$ states and magnetic field strengths $B$. The optimized variable $E_{\textrm{fs}}=(E_1,E_2,E_3)$ is the bare fine-structure splitting of the $5\triplet D$ levels. In the argument below we assume only the formalism of atomic structure theory and the data from our experiment. To determine the bare $5\triplet D$ transition energies from our data, consider the Hamiltonian
\begin{equation}
    \hat{H}(B) = \hat{H}_{\textrm{fs}} - B\hat{\mu}_z,
    \label{eqn:hamiltonian}
\end{equation}

where $\hat{\mu}_z = \mu_B(\hat{L}_z + g_s \hat{S}_z)/\hbar$ is the coupling of the orbital and spin angular momenta of the electron with a magnetic field of strength B pointing in the $z$-direction, $\mu_B$ is the Bohr magneton and $g_s$ is the electron spin $g$-factor. The fine structure Hamiltonian $\hat{H}_{\textrm{fs}}$ is diagonal in the $|LSJ m_J\rangle$ basis with eigenvalues $E_{\textrm{fs}}$,
\begin{equation}
  \hat{H}_{\textrm{fs}}|LSJm_J\rangle = E_{\textrm{fs},LSJ}|LSJm_J\rangle,
\end{equation}
which are degenerate for all $m_J$ with fixed $J$. The magnetic moment $\hat{\mu}_z$ couples states of different $J$, and is instead diagonal in the $|L m_L S m_S\rangle$ basis. In the $|LSJ m_J\rangle$ basis the matrix elements of $\hat{H}(B)$ are, with abbreviated notation,
\begin{equation}
\begin{aligned}
H_{J',J} =& \langle J'|\hat{H}|J \rangle\\
  =& E_{\textrm{fs},J} - B \frac{\mu_B}{\hbar} \sum_{m_L} (2m_J - m_L)C_{J,'m_L}C_{J,m_L},\\
\end{aligned}
\end{equation}
where $C_{J,m_L} = \langle LSJ m_J|L m_L S m_S\rangle$ is shorthand for the Clebsch-Gordan coefficients. For $B>0$, the contribution of $\hat{\mu}_z$ breaks the degeneracy of $\hat{H}_{\textrm{fs}}$, giving rise to the Zeeman shift. 

\begin{figure}[t]
    \centering
    \includegraphics[width=0.48\textwidth]{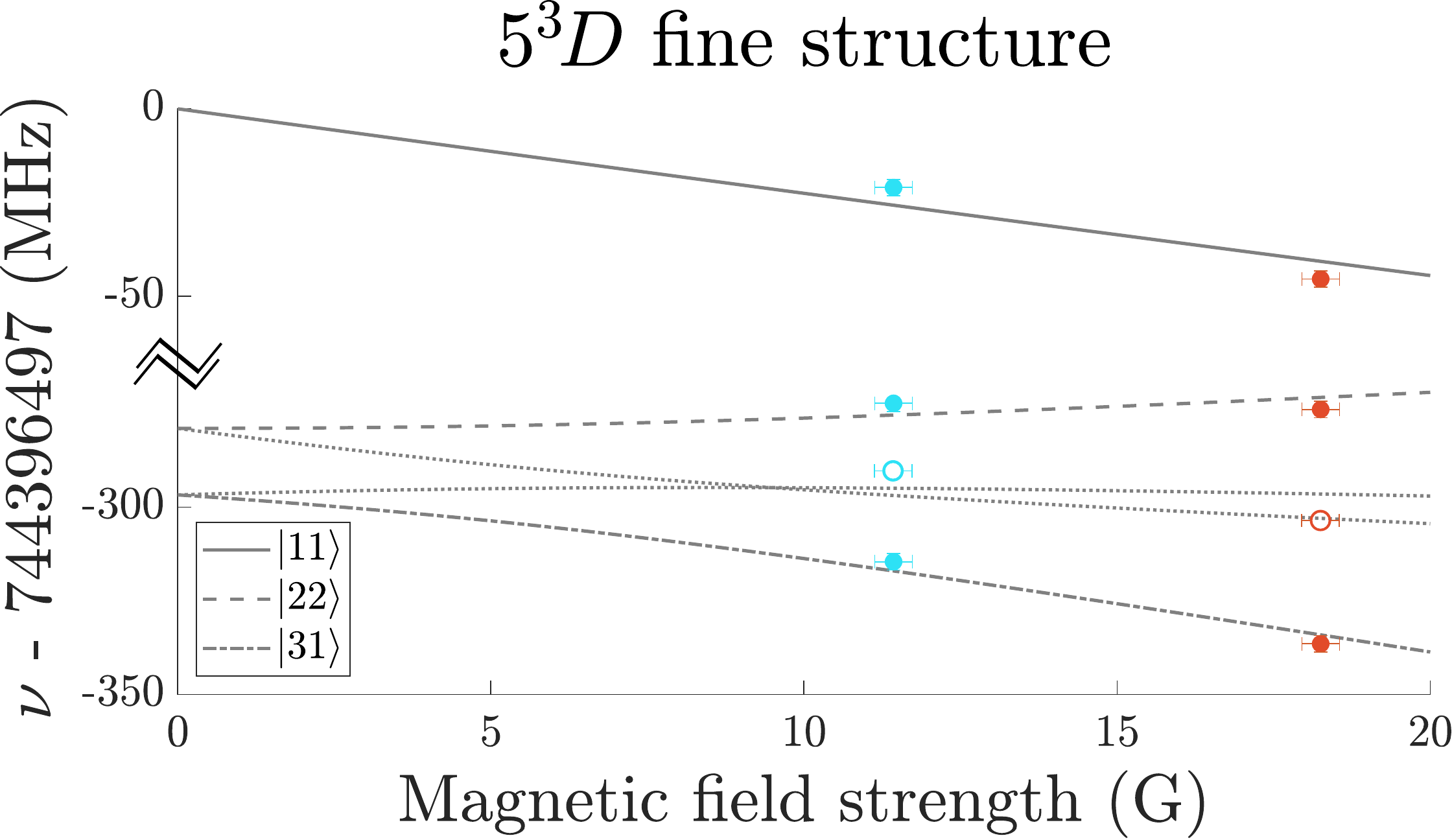}
    \caption{Determining the $5\triplet D$ fine-structure splitting. The values for the $|J,m_J\rangle=5\triplet D_J(m_J)$ levels (grey lines) at $B=0$ are fixed by solving the optimization problem (Eqn. \ref{eqn:opt-problem}), constrained by the fitted peak centres (filled circles). The saturated peaks are not used to constrain the levels, but are shown with hollow circles and the corresponding frequencies predicted by our method are shown in dotted lines.}
    \label{fig:fitting_3D}
\end{figure}

\begin{table*}[t]
\caption{Summary of results for each transition. After correcting for the AOM and vapor cell shifts we extract the centre frequencies from Lorentzian fits with statistical error at the $10\textrm{kHz}$ level. We obtain the field-free energies after correcting for Zeeman shifts, shown with theoretical predictions in the row below. We show the difference between our measurements and theoretical predictions for direct comparison. Observed full width at half maximum line widths (FWHM) of the Lorentzian fit to each line are shown in comparison to predicted linewidths as given in \cite{Drake07}. All values are in MHz with uncertainty in the final digit in parentheses.}
        \begin{tabular}{c c c c c c c c c c c}
      \hline\hline
      Transition                        & &  Frequency $f_\textrm{exp}$ & &  Frequency $f_\textrm{theory}$ & Ref. &  $f_\textrm{exp}-f_\textrm{theory}$      & &  $\textrm{FWHM}_{\textrm{exp}}$  & &  $\textrm{FWHM}_{{\textrm{pred}}}$ \\
      \hline
        $2\triplet P_2 - 5^3\mathrm{S}_1$ & &  727,303,250(2) & &  727,303,244.6(4)   & \cite{Drake07} &    5(2)    & &  3.4(5)  & &  1.5\\ 
        $2\triplet P_2 - 5^3\mathrm{D}_1$ & &  744,396,497(7) & &  744,396,511.1(7)   & \cite{Yerokhin20} &  -14(7)    & &  5.8(6)  & &  2.6\\
        $2\triplet P_2 - 5^3\mathrm{D}_2$ & &  744,396,218(7) & &  744,396,227.6(7)   & \cite{Yerokhin20} &  -10(7)    & &  4.2(5)  & &  2.6\\
        $2\triplet P_2 - 5^3\mathrm{D}_3$ & &  744,396,200(7) & &  744,396,208.3(7)   & \cite{Yerokhin20} &   -8(7)    & &  4.0(1)  & &  2.6\\
        $2\triplet P_2 - 5^1\mathrm{D}_2$ & &  744,430,347(7) & &  744,430,343.1(7)   & \cite{Yerokhin20} &    4(7)    & &  3.2(1)  & &  2.2\\  
      \hline\hline
    \end{tabular}
  
  \label{tab:results}
\end{table*}
The solution of Eqn. \ref{eqn:opt-problem} is illustrated in Fig. \ref{fig:fitting_3D}. If we define the energies $E_{\textrm{fs}}$ relative to the $2\triplet P_2(m_J=2)$ state, then we can read the predicted transition frequencies directly from the eigenvalues of $\hat{H}$ via $\nu_{J,m_J}^{\textrm{pred}}=E_{J,mJ}(B)/h$, where $h$ is Planck's constant. The observed frequencies $\nu_{J,m_J,B}^{\textrm{obs}}$ used in this procedure exclude the saturated peaks because their centre frequencies cannot be determined accurately. The triple $E_{\textrm{fs}}$ which minimizes the cost function (Eqn. \ref{eqn:opt-problem}) corresponds to the $2\triplet P_2 - 5\triplet D_J$ intervals at $B=0$, as listed in Tab. \ref{tab:results}. Again, the difference in our determination of the field-free splitting is consistent within 2$\sigma$ predictions in \cite{Drake07} after accounting for systematic effects (Tab. \ref{tab:errors}). 

\section{Shifts, broadening, and errors}

\begin{table}[b]
\caption{Error budget for the determination of the peak centre frequencies.  The master laser for our pump beam is described in \cite{Shin16}. AOM stabilities were checked with an RF spectrum analyser. See \cite{Thomas20} for measurement of the Cesium cell shift and probe beam lock drift. The shift and uncertainty from the Zeeman shift (ZS) varies between the lines, so these contributions are omitted from the total. All values are in MHz.}
  \label{tab:errors}
  \begin{tabular}{c c c}
      \hline\hline
          Source & Shift & Broadening  \\
      \hline
          Wavemeter ($5\triplet S_1$)& 0(1.3) & - \\
          Wavemeter (all other lines)& 0(6.7) & - \\
          Pump lock & - & 4$\times10^{-2}$ \\
          Pump AOM & - & 0.3 \\
          Probe lock & - & 0.3\\
          Probe AOM & -189 & 1$\times10^{-6}$\\
          Zeeman & Variable & Variable \\
          Recoil & - & 1.4$\times 10^{-3}$ \\ 
          Doppler & - & 1.6 \\
          Cs cell & -1.9 & 0.4 \\
          \textbf{Total} ($5\triplet S_1$ level) & -190.9(1.7)+ZS& 2.2\\
          \textbf{Total} (all other levels) & -190.9(6.7)+ZS& 2.2\\
      \hline\hline
  \end{tabular}
  
\end{table}


The results of our measurements are reported in Tab. \ref{tab:results} and are consistent with theoretical predictions \cite{Morton06} within $2\sigma$ or less. The accuracy of our determinations of the field-free transition energies is limited by the absolute accuracy of the wavemeter. High Finesse specifies \cite{HighFinesseDoc} a 3$\sigma$ accuracy of 2MHz within 2nm of a calibration line (as in the transition to the $5\triplet S_1$ state), and 10MHz for all other lines measured in this work. Because we use the wavemeter to lock the seed light for the doubler, the uncertainty is doubled in determinations of the absolute probe frequency. We adopt the corresponding 1$\sigma$ accuracy in order to be consistent with other terms in our error budget, as displayed in Table \ref{tab:errors}. We note that this specification does not depend on the specific difference between the calibration and measured wavelengths, and may vary due to nonlinear dispersion of the wavemeter optics. Without an independent calibration we cannot rigorously constrain this source of error, which would be overcome with the instrumental improvements discussed below. As such, we state the 1-sigma errors determined in this way with the caveat that they may be slightly underestimated. Still, all measured frequencies are consistent with predictions to within 2.5$\sigma$. Finally, we note that the the $5\triplet S_1$ transition line is 1.9nm away from the calibration line, and as such the 2MHz uncertainty may again be a slight underestimate.

The linewidth of the pump and probe laser sources are 40kHz \cite{Shin16} and 200kHz \cite{Thomas20}, respectively. The laser lock error has a standard deviation of $100$kHz. The additional contribution from the pump and probe AOMs are 300kHz and 1Hz, respectively, as determined with an RF spectrum analyser.

Kinetic effects do not contribute any significant uncertainty in our frequency measurements, rather they just broaden the observed peaks. The pump light is applied by two counterpropagating beams, subtending angles of 15$^\circ$ and 195$^\circ$ relative to the direction of propagation of the probe beam. Photon absorption events contribute a recoil velocity of magnitude $\cos(15^\circ)\cdot\hbar k/m\approx6$mm/s, imparting a Doppler shift of order 1.4kHz. Atoms may absorb probe light after absorbing a photon from the pump beams, but not after decaying again, so the decay events do not contribute. Because there are two counterpropagating pump beams, the resulting contribution is a negligible broadening, especially in comparison with the thermal broadening. For a sample of $^4$He at the Doppler limit of 38$\mu$K, the Doppler broadening of the pump and probe beams due to the temperature of the cloud are 600kHz and 1.6MHz, respectively. The difference between predicted and observed line widths is well accounted for by these broadening effects.

\begin{figure}[b]
\includegraphics[width=0.48\textwidth]{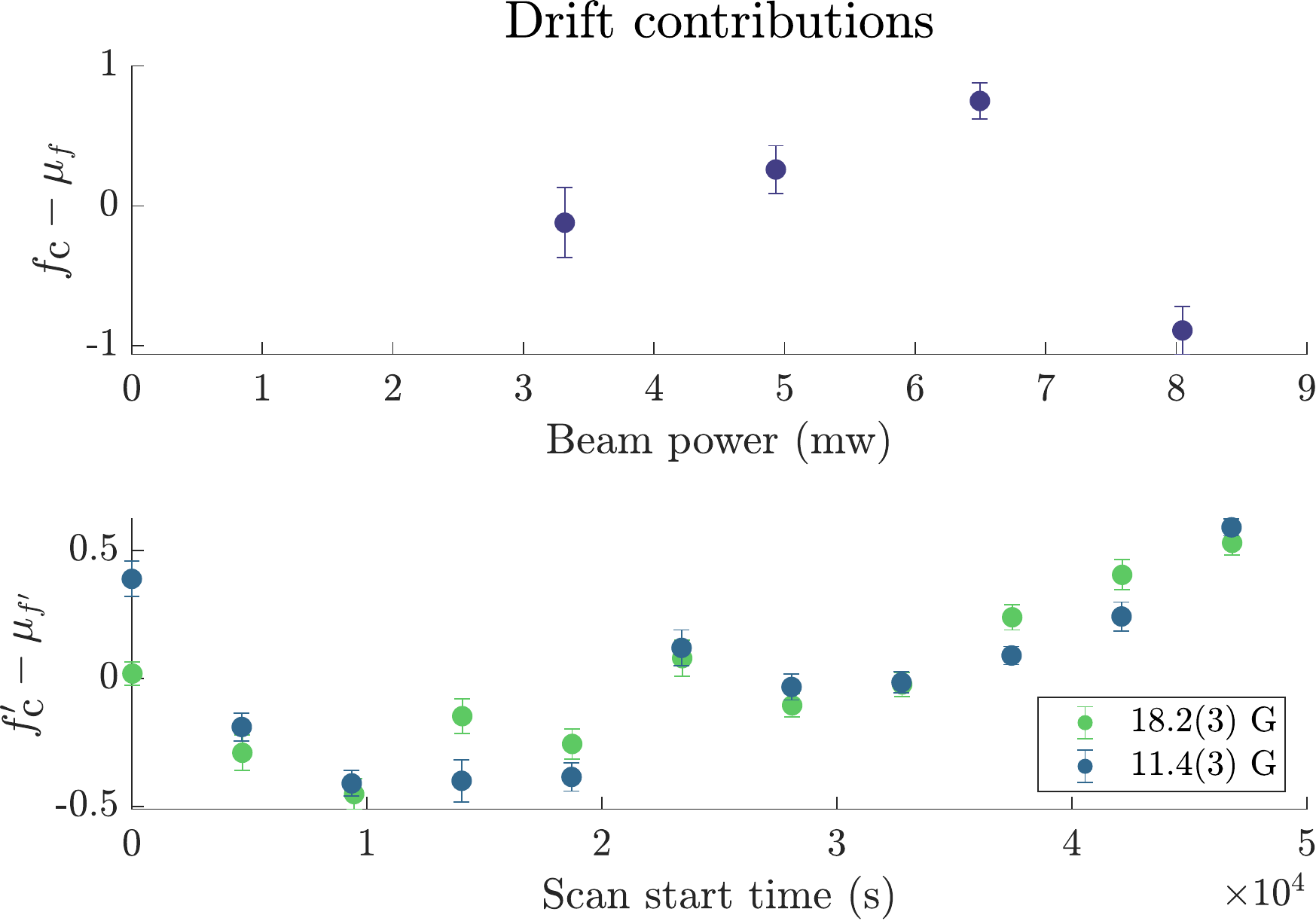}
\caption{Top: Variation in fitted centre frequency for single scans across the $5\triplet D_1$ line versus applied laser power. The measurements at increasing beam power were not taken in chronological order Bottom: Variation in fit center frequency for the $2\triplet P_2 - 5\singlet D_2$ between scans. The value of the fitted peak centre $f_\textrm{c}$ is shown for each field strength, relative to the mean $\mu$ of all values for that field strength.}
\label{fig:power_drift_combined}
\end{figure}

We determine no significant contribution from the AC Stark effect. During measurements of the $5\triplet D_1$ transition with varying probe beam powers, we found that any dependence of the centre frequency on the laser power was dominated by the drift in the wavemeter output, as shown in Fig. \ref{fig:power_drift_combined}. For the triplet-singlet transition, the increase in laser intensity is more than compensated for by the reduced dipole matrix element, and hence we come to the same conclusion.

\section{Discussion}

We performed multilevel laser absorption spectroscopy of excited state transitions in ultracold helium. This includes the first observation (to our knowledge) of the forbidden $2\triplet P_2 - 5\singlet D_2$ transition. Our measurements agree with current predictions within our error budget and suggest that the $93\sigma$ difference between previous measurements \cite{Martin60} and predictions \cite{Morton06} of the $\PStateManifold_2  -  \UpperS$ and $\PStateManifold_2  -  \UpperStateManifold$ intervals are due to an unkown systematic error \footnote{As a historical note on the advancement of experimental methods, Martin's measurements were made using a nitrogen-cooled helium discharge lamp fed through an in-vacuum prism onto photographic plates, on which the line separations were measured by hand with a ruler.}. 

The techniques described here are readily extensible to other opportunities in $^4$He structure measurements. For example, while there is an outstanding 7.4$\sigma$ disagreement between the predicted and observed singlet-triplet interval for the n=3 level in $^3$He \cite{Morton06,Derouard80}, the corresponding transition in $^4$He has never been directly measured. An indirect measurement in $^4$He could be made with the techniques described here by taking the difference betweeen the $2\triplet P_2 - 3\triplet D_2$ and $2\triplet P_2 - 3\singlet D_2$ transitions near 587.6nm and 587.4nm. While the latter transition is also spin-forbidden, it is predicted to be an order of magnitude stronger than the $2\triplet P_2 - 5\singlet D_2$ transition reported here \cite{Morton06}.

Further, energies of other $2L-nD$ transitions in $^4$He are a few MHz larger than predicted, apparently independent of $L$ \cite{Wienczek19,Yerokhin20}. Our results break this trend, and invite independent verification. Further study of transitions between states from different shells to MHz precision or better, in particular the prospective study of the $2\triplet P - 3 D$ intervals, would also provide an independent constraint for this disagreement.

Simply exchanging the light source would suffice to make these measurements, but a definitive comparison with theory would require an improved frequency reference. For instance, the hypothetical $10/n^3$ MHz shift could be checked by a measurement of the $2~P-n~D$ transitions accurate to sub-MHz precision. The associated theoretical uncertainties are about this size, dominated by the 700 kHz uncertainty in the lower state \cite{Pachucki17,Wienczek19}. As the $\alpha^7$ terms could improve the theoretical accuracy to as little as 10kHz \cite{Pachucki17}, this more challenging precision appears to be a more appropriate budget, and readily achievable with current methods. Reference-locked optical frequency combs can readily achieve kHz accuracy or better \cite{Luo15,Rengelink18}. Magnetic field strengths can be determined by RF spectroscopy with sub-kHz accuracy and so would not present a serious limitation. Improving the AOM frequency stability would provide sufficient accuracy to account for laser-induced stark shifts, likely leaving systematic drifts as the dominant source of error.

Extending these methods to direct measurements on $^3$He would also permit isotope shift measurements from forbidden excited-state transitions in $\triplet$He. Theoretical calculations of isotope shifts are already accurate to the sub-kHz level, so such measurements would be even more demanding than the prospects above. Existing demonstrations of comparable accuracy \cite{Rengelink18} show such measurements are worthy challenges whose completion can access femtoscale nuclear structure information via optical atomic spectroscopy.

\section{Acknowledgements}

We thank Gordon Drake for informative discussions. We are especially grateful to Chris Vale and Sascha Hoinka for the generous loan of the light source, and to Carlos Kuhn for technical assistance in setting up for a series of measurements including this work. The project  was supported by Australian Research Council (ARC) Discovery Project Grants DP160102337 and DP180101093, and Linkage Project LE180100142. J.A.R. and K.F.T. are supported by ANU Postgraduate Research and Australian Government Research Training Program (RTP) Scholarships, respectively.

\bibliography{bibliography}

\end{document}